\documentstyle[onecolumn]{mn}
\def\mo{m_{\odot}}
\def\al{\alpha}
\def\cmo{M_{\odot}}
\def\be{\begin{equation}}
\def\ee{\end{equation}}
\def\cen{$\omega Cen~$}
\def\tuc{47 Tuc }
\def\n6{NGC 6397}
\def\ltorder{\hbox{ \rlap{\raise 0.425ex\hbox{$<$}}\lower
0.65ex\hbox{$\sim$} }} 
\begin{document}
\title{On the effects of dynamical evolution on the initial mass function of
globular clusters}

\author[E.Vesperini,
D.C.Heggie]{E.Vesperini\thanks{Present address: Department of Physics
and Astronomy, University of Massachusetts, Amherst, MA 01003-4525,
USA;  E-mail:vesperin@falcon.phast.umass.edu},
D.C.Heggie\thanks{E-mail:d.c.heggie@ed.ac.uk}\\ Department 
of Mathematics and Statistics, University of Edinburgh, King's
Buildings, Edinburgh, EH9 3JZ}
\maketitle
\begin{abstract}
 In this paper we show the results of a large set of $N$-body
 simulations modelling the evolution of globular clusters
 driven by relaxation, stellar evolution, disk shocking and
 including the effects of the tidal field of the
 Galaxy. We investigate the evolution of multi-mass models with a
 power-law initial mass function (IMF) starting
 with different initial masses, concentrations, slopes of the
 IMF and located at different galactocentric distances.
 We show to what extent
  the effects of the  various evolutionary processes alter
 the shape of the IMF and to what extent these
 changes depend on the position of the cluster in the Galaxy. Both
 the changes in the global mass function and in the local one
 (measured at different distances from the cluster center) are
 investigated showing whether and where the local mass
 function keeps memory of the IMF and where it provides a good
 indication of the current global mass function. 

The evolution of the population of white dwarfs is also followed in
detail and we supply an estimate of the fraction of the current
value of the total mass expected to be in white dwarfs depending on
the main initial conditions for the cluster (mass and position in the
Galaxy). Simple analytical expression by which
it is possible
to calculate  the main quantities of interest (total mass, fraction of
white dwarfs, slope of the mass function) at any time $t$ for a
larger number of different initial conditions than those investigated
numerically have been derived.
\end{abstract}
\begin{keywords}
globular clusters:general -- stellar dynamics
\end{keywords}
\section{Introduction}
Investigation of the mass function of  globular clusters is of great
importance for a variety of problems in astrophysics covering  star
formation processes, the dynamical evolution of stellar systems and
the nature of dark matter in the Galaxy. Large progress has been made
in recent years both by ground based observation and, more recently,
thanks to observations by HST. Nevertheless most
of issues
concerning the shape of the initial mass function (IMF), its
dependence on cluster parameters, the actual relevance of dynamical
processes in its evolution and the relation between the IMF and the
present-day mass function (PDMF) are still matters of debate.

The first investigation addressing the dependence of the slope of the
mass function on cluster structural parameters and metallicity was
carried out by McClure et al. (1986) who found  the slope of the
PDMF for a sample of six galactic clusters to be correlated with their
metallicity, the low-metallicity clusters having  steeper mass
functions.
In subsequent work Capaccioli, Ortolani \& Piotto (1991), Piotto
(1991) and  Capaccioli, Piotto \& Stiavelli (1993) have considered a
larger sample of  
clusters  and have questioned the conclusion of McClure et al.
and showed the correlation between the slope of the PDMF and  the
position of the cluster in the Galaxy to be stronger than that with
the metallicity. Finally Djorgovski, Piotto \& Capaccioli (1993) have
addressed this 
problem again by multivariate statistical methods and have
concluded that both the position in the Galaxy (galactocentric
distance and height above the disk) and the metallicity play some role
in determining the slope of the PDMF but  the former is more
important than the latter. 
The observed correlation is in the sense of
clusters more distant from the galactic center or from the disk having
steeper mass functions.

The data used in the above works are from ground based observations and
the slopes are measured for a limited range of star masses
($0.5<m/m_{\odot}<0.8$). Recent investigations of
the luminosity function of some galactic globular clusters 
by HST have
been able to extend the available data to fainter magnitudes (Paresce,
Demarchi \& Romaniello 1995, De Marchi
 \& Paresce 1995ab, Elson et al. 1995, Piotto, Cool \& King 1996,1997,
Santiago, Elson \& Gilmore 1996). HST data for
 for \tuc, \cen, \n6, M15 and M30 are now available.
These clusters span a wide range of values of  metallicity, 
their structural parameters suggest they have undergone a very different
dynamical evolution and the issue concerning the origin of
 the shape of the PDMF has been addressed again in the
light of this new data.

De Marchi \& Paresce  (1995b) compare the MF of \n6, M15 and \tuc showing
that all these clusters have a flat MF for low-masses; they point out
that the MF is flat at the low-mass end for both  \n6 and M15 and that these
MFs are very similar though these clusters are
likely to have had a different dynamical history. As for \tuc,
this is shown to have a different MF from M15 and \n6.
Noting that the metallicity of \tuc is very different from
that of \n6 and M15 De Marchi \& Paresce 
make the hypothesis that the differences
between the MFs of these clusters might be due to a different initial
mass function (IMF) depending on the metallicity, thus giving new
support to the conclusion of McClure et al. (1986), with the
subsequent dynamical evolution playing no relevant role.

However in a recent  work,  Santiago et al. (1996) show that the MF
of \cen, whose metallicity is similar to that of \tuc, is steeper than
the MF of \tuc, and cast some doubt on  the scenario supported by
De Marchi and Paresce. Santiago et al. 
point out that if one assumes a universal IMF, the comparison of the
MF of \cen with those of \n6, M15, \tuc would indicate that the latter
clusters have experienced  significant
dynamical evolution with  strong depletion of low-mass stars.
Finally Piotto et al. (1996,1997) argue that the reason why De Marchi and
Paresce get a similar MF for \n6 and M15 is that they compare only the
low-mass end and show that, by comparing the LF including the
 data for the bright end, 
\n6 appears to be markedly deficient in
faint stars. As the metallicities of \n6 and M15 are very similar, this
result lends strong support to the hypothesis that the MF of \n6 is
flattened by dynamical processes. King (1996) notes that this
hypothesis  is further enforced
by the comparison of the orbits of \n6 and M15, as obtained by
Dauphole et al. (1996); according to this work \n6 would be more
affected by the tidal interaction with the Galaxy as it would cross the
disk more frequently and would have a smaller perigalactic distance
than M15.

Additional observations covering a larger range of cluster
parameters  are  necessary, as
well as theoretical investigations addressing both the problems
connected with the stellar evolution models (see Alexander et al.
1997, D'Antona \& Mazzitelli 1996 for two recent works in this
direction) allowing a better 
determination of the mass-luminosity relationship for low-mass
low-metallicity stars (see e.g. Elson et al. 1995 for a clear
presentation 
of the problems due to the uncertainties on $M-L$ relation)
and those connected with the dynamical evolution, thus clarifying the
efficiency of evolutionary processes in modifying the IMF.

As for this latter aspect the situation is far from being clear:
simple semi-analytical models by Stiavelli et al. (1991), Stiavelli,
Piotto \& Capaccioli (1992) and
Capaccioli et al. (1993) suggest that
disk shocking could play a relevant role in establishing the observed
correlation between the slope of the PDMF and the position in the
Galaxy and some indications on the role of evaporation due to two-body
relaxation come from many multi-mass Fokker-Planck
investigations of the dynamical evolution of clusters (see e.g.
Chernoff \& Weinberg 1990, Weinberg 1994) but no firm conclusion has
been reached on
the relevance of various dynamical processes in giving rise to the
observed correlation between the slope of the mass function and
position in the Galaxy  and the interplay between different dynamical
processes has not been fully explored.

In this paper we show the results  of a set of $N$-
body simulations including the effects of the presence of the tidal
field of the Galaxy, stellar evolution, disk
shocking, two-body relaxation and spanning a
range of different initial conditions for the
 mass and concentration of the cluster, the slope of the IMF and
the distance from the galactic center.

The main goal of our theoretical investigation is that of assessing
the importance of various evolutionary processes in altering 
 the mass function of a globular cluster, trying to establish to
what extent these processes can be  responsible for the  differences
observed between the  MFs in galactic globular clusters.
Having included the effects of stellar evolution in our investigation
we will be able to address some other issues concerning the evolution
of the stellar content of globular clusters. In particular we will
focus our attention on the fraction of white dwarfs expected to be
present in a cluster during
the different stages of its dynamical evolution. These are of
increasing interest for a variety of reasons. It had long been
realised that white dwarfs should exist in some abundance in globular
clusters, and they have been included routinely in dynamical
models. Otherwise, however, little attention was paid to them while
they remained unobservable. That situation has changed dramatically in
recent years (e.g. Richer et al. 1995), and it is now claimed that
white dwarfs formed up to 9 Gyr ago have been detected (Richer et
al. 1997). Furthermore recent suggestions have been made that white
dwarfs could account for as much as 50 percent of the total mass of
some clusters (Heggie \& Hut 1996) or up to 85 percent in some
locations in other clusters (Gebhardt et al. 1997). This sudden burst
of interest in white dwarfs will stimulate renewed attention in their
influence on cluster dynamics, and has motivated the work reported in
section 3.3.

In Sect.2 we describe the method used for
our investigation. In Sect.3 the results of simulations including the
effects of stellar evolution and two-body relaxation only are
described, and some analytical expressions are derived for the main
quantities of interest for this work. In Sect.4  the results of runs
including also the effects of disk shocking are described and, where
possible, we generalize the analytical expressions obtained in
sect.3 . The dependence of the results on the initial number of
particles used in the simulations are discussed in sect.5. Summary and
conclusions are in Sect.6.
\section{Method}
All the $N$-body simulations made for our investigation have
been carried out using a  version of the code NBODY4 (Aarseth
1985) including mass loss of single stars due to stellar evolution,
the effects of disk shocking and of the tidal field of the Galaxy.
The code is a direct summation code adopting an Hermite integration
scheme and a binary hierarchy of time-steps; a GRAPE-4 board
containing 48 HARP chips (Makino, Kokubo \& Taiji 1993) connected
to a  Dec Alphastation 3000/700 has been used for  the evaluation of
the forces and force derivatives required by the Hermite scheme.
Most of the simulations carried out in this work start with a total
number of particles $N=4096$ and it took  a CPU time ranging from
about 3
to 15 hours for each of them to be completed depending on the initial
concentration of the system and its galactocentric distance (for 
further information on the  code and the performance of the board 
see e.g. Aarseth 1994, Aarseth \& Heggie 1997a).

In our simulations we have assumed the cluster to be on a circular
orbit and to move in a Keplerian potential determined by a point mass
$M_g$ equal to the total mass of the Galaxy inside the adopted
galactocentric distance $R_g$; the value of the circular speed has
been taken equal to $v_c=220$ km/s. If we denote the angular velocity
of the cluster around the galaxy by $\omega$ and take the coordinates of
a star relative to the center of the cluster with the $x$-axis
pointing in the direction away from the galactic center and the y-axis
in the direction of the cluster motion   we can write the equations of
motion for a star in the cluster as (see e.g. Chandrasekhar 1942)
\begin{eqnarray}
&& {d^2 x \over dt^2}-2\omega {d y\over d t}-3\omega^2 x=F_x\nonumber
\\
&& {d^2 y \over dt^2}+2\omega {d x\over d t}=F_y
\\
&& {d^2 z \over dt^2}+\omega^2 z=F_z\nonumber\\
\nonumber 
\end{eqnarray}
where $\bf {F}$ represent the acceleration from other stars in the cluster
and the terms involving $\omega$ on the left-side of the equations
are due to the 
Coriolis, centrifugal and tidal field accelerations.

King's models (King 1966) with different concentrations have been used
to produce initial conditions  and the standard $N$-
body units (Heggie \& Mathieu 1986) (total mass $M=1$, $G=1$ and
initial energy equal to -1/4) have been adopted. The value of $\omega$
in $N$-body units is given by its initial relationship with the tidal radius
(see Aarseth \& Heggie 1997a)
\be
r_t^*=(3\omega^{*2})^{-1/3},
\ee
where the $*$ denotes quantities in $N$-body units.

Masses of stars have been assigned according to a power-law mass
function
$$
dN(m)=m^{-\alpha}dm
$$
between $0.1 \mo$ and $15 \mo$ and initially there is no equipartition
of energies of stars with different masses.

Disk shocking  has been included according to the
model described in Chernoff, Kochanek \& Shapiro (1986). Following
Spitzer (1958), Chernoff et al. describe the motion of a  single star
in the cluster as
that of an harmonic oscillator with a frequency $\Omega$ perturbed by
the force due to the disk. Assuming  the orbit to be perpendicular to
the plane of the disk and choosing the direction of motion along the
$y$-axis, the equation of motion of a star in a frame at rest with
respect to the 
center of the cluster is
\be
{\hbox{d}^2 \delta y \over \hbox{d} t^2}+\Omega^2 \delta y=g(t)\delta y
\ee
where $\delta y$ is the distance between the star and the center of the
cluster, $y$ the height of the center of the cluster over the disk and
$g(t)$ is the differential acceleration due to the disk.

Modelling the disk as a one-component isothermal self-gravitating gas and assuming
an exponential decay for the disk surface density, the
acceleration due to the disk is given by
\be
K_y=K_0 \hbox{tanh} (y/y_0)\hbox{e}^{-R_g/h}.
\ee
where $R_g$ is the galactocentric distance and $y_0$ and $h$ are the
disk scale height and characteristic scale length respectively.

Starting from the above equations Chernoff et al. show that 
 the change of velocity in the $y$ direction suffered
by a star due to the disk shocking can be written as 
\be
\Delta v_y=\delta y I_c(\Omega) \label{dvds}
\ee
where $I_c$ is
\be
I_c={\pi \Omega \over 2 \nu^2 \sinh(\pi \Omega/2\nu)}{K_0 \over
z_0}\hbox{e}^{-R_g/h}
\ee
and $\nu=v_c/2y_0$ is the frequency of disk crossing.
When $\nu\gg \Omega$, the above expression provides the same result
which would be obtained in the impulse approximation (see e.g. Spitzer
1987) while for $\nu <\Omega$ the crossing is adiabatic and the change
in star's velocity tends exponentially to zero.
Notice that, in the pure impulse approximation, (as adopted, for
example, by Capaccioli et al. (1993)), the effect of a disk shock is 
minimised by choosing an orbit which crosses the disk
perpendicularly, as we have done, because then $\nu$ is maximised
(for a given circular speed). The adiabatic correction, however,
suppresses low-frequency disk shocking, and the net result is that
the effect of disk shocking is almost independent of the inclination
 of the orbit over a wide range (Chernoff, Kochanek \& Shapiro 1986).

In our investigation we have adopted the two-component disk model
obtained by Chernoff et al. from fitting  Bahcall's (1984) determination
of acceleration in the solar neighbourhood
\be
K_y(R_g=8~\hbox{Kpc},y)=\sum_i K_{0i}\tanh(y/y_{0i})
\ee
with
$$
K_{01}=3.47\times 10^{-9} \hbox{cm s}^{-2};~~y_{01}=175 \hbox{pc}
$$
$$
K_{02}=3.96\times 10^{-9} \hbox{cm s}^{-2};~~y_{02}=550 \hbox{pc};
$$
and the scale length has been taken equal to $h=3.5$ Kpc (see e.g.
Bahcall, Schmidt \& Soneira 1982).

In the  simulations  including the effects of disk shocking, each
half-orbital period of the cluster around the Galaxy the $y$-component
of the velocities of
all the stars in the system have been changed according to
eq.(\ref{dvds}) (where $I_c$ has been replaced by the sum, $I_{c1}+I_{c2}$,
with $I_{c1,2}$ being the function $I_c$ evaluated for the two  disk
components) with the frequency $\Omega$ taken to be equal to the ratio
of the current value of tangential velocity of the star to its radial
distance from the cluster center.

Stellar evolution has been modelled by assuming that the mass lost by
each star immediately escapes from the cluster (this is likely to be a
reasonable approximation since the escape velocity from a typical
globular cluster is of the order of 10 km s$^{-1}$) and thus the mass of each
star has been decreased at the appropriate time by an amount depending
on the initial value of the mass itself. 
The fraction of mass lost by stellar evolution and the time when mass
loss has to take place in the simulation have been calculated adopting
the same model used in Chernoff \& Weinberg (1990) (see also Fukushige
\& Heggie 1995): stars whose initial mass is larger than $8~ \mo$ end
their life as neutron stars with a mass equal to $1.4~ \mo$, stars with
$4.7<m/\mo<8$ are assumed to leave no remnant and stars less massive
than $4.7 \mo$ produce white dwarfs with a mass equal to
$0.58+0.22(m/\mo-1)$; the times the mass must be removed are
determined by a linear interpolation of the main sequence times
calculated by Iben \& Renzini (1983) and 
reported in Table 1B of Chernoff \& Weinberg (1990). We have assumed
that there is no kick velocity
from the supernova explosion for stars ending their evolution as
neutron stars. Though consistent with the assumptions of Chernoff \&
Weinberg (1990)  this assumption is very likely to be incorrect
(Drukier 1996). Nevertheless, a more realistic treatment would make
almost no difference to our results, as the mass fraction in neutron
stars rarely exceeds 1 percent for our assumed IMF (Fig.1d).

The last issue to be addressed concerns the scaling from time in
$N$-body units to astrophysical units; as we want to include the
effects of disk shocking and stellar evolution,
an appropriate time scaling is 
necessary not only for a correct application of $N$-body results to
real clusters but also for a proper determination of
the times when the mass of stars must be removed due to stellar
evolution and the change in the velocities of stars due to disk
shocking have to be made.
An extensive investigation of this issue has been carried out by
Aarseth \& Heggie (1997a) (see also Fukushige \& Heggie 1995) who have
shown that, depending on the time 
scale of the physical process (stellar evolution or relaxation)
determining the evolution of the cluster, the scaling must ensure that
the 
ratio of  ``real'' time to $N$-body time must be equal to the ratio of
``real'' to $N$-body crossing or relaxation time.

In fact, as discussed in Aarseth \& Heggie (1997a), if the cluster lifetime is
shorter than the time scale in which relaxation effects become
important (e.g. because of strong mass loss due to stellar evolution
causing the cluster to disrupt) the proper factor to scale $N$-body
time to real time is given by the ratio of the crossing time of
$N$-body system to that of the real system. On the other hand if the
cluster survives for a time long enough to be affected by relaxation the proper
conversion factor is provided by the ratio of relaxation times of
$N$-body and real system thus  ensuring that the number of 
relaxation times elapsed is the same in the $N$-body system and in the
real cluster.
Possibly a variable time scaling, taking into account the transition
from a phase dominated by stellar evolution effects to one dominated
by relaxation, might be the best choice (see Aarseth \& Heggie 1997a
for a detailed discussion of this point).

Since all the systems considered in our work do not disrupt quickly
due to the effects of stellar evolution but survive for a lapse of time
during which relaxation effects become important, we have scaled time
by the ratio of relaxation times (half-mass relaxation times have been
used) of $N$-body to real system.

\section{Results: two-body relaxation and stellar evolution}
\subsection{Global mass function}
A set of simulations  not including disk shocking have been  run to 
establish to what extent differential escape due to two-body
relaxation (and mass loss due to stellar evolution) can alter
the mass function of a globular cluster. 
Since the evaporation rate due to
two-body relaxation is larger for low-mass stars than it is for the
high mass ones (see e.g. Spitzer 1987, Giersz \& Heggie 1997 for some
recent $N$-body
simulations where this is clearly shown) a flattening of the IMF as dynamical
evolution of a cluster proceeds is expected. This process could also
be  responsible for the  observed correlation between the slope of the
mass function and the galactocentric distance; in fact, as the evaporation
rate is inversely proportional to the relaxation time, the largest
changes in the IMF, for a fixed value of the cluster mass, are to take
place closer to the galactic center where the size of clusters are
smaller and the relaxation time shorter. 

Table 1 summarizes the initial conditions and the main results 
for the set of runs done.
The parameter $F_{cw}$ is defined as
\be
F_{cw} \equiv {M_i \over \cmo}{ R_g\over \hbox{Kpc}}{1 \over
\ln(N)}{220 \hbox{km s}^{-1}\over v_c}
\ee
where $M_i$ is the initial mass of the cluster, $N$
the total initial number of stars, $R_g$ the distance from the
galactic center and $v_c$ the circular velocity around the Galaxy. 
This parameter,
introduced by Chernoff \& Weinberg (1990), is proportional to the
relaxation time and  thus clusters having the
same value of $F_{cw}$ form a family of models evolving in the same
way provided that relaxation dominates (obviously this is true for clusters all with the same initial
concentration and IMF).
Both the initial concentration and the slope of the IMF have been
varied in order to investigate the dependence of the results on these
quantities.  The slope of the mass
function, at $t=15$ Gyr, reported in Table 1 is that measured for
main sequence 
stars with $0.1 <m/m_{\odot}<0.5$. 

It is evident that
evaporation through the tidal boundary due to two-body relaxation
gives rise to a significant trend between the slope of the MF and the
distance from  the galactic center. As the results of runs starting
with $W_0=5$ show, the trend established does not depend significantly
on the initial concentration. On the other 
hand, the slope of the IMF seems to have a more important effect on
the formation of this correlation, and more in general on the extent
the IMF evolves, as shown by the results of the runs
with $\alpha_{i}=3.5$. In this case the differences between the initial
and the final value of $\alpha$ are much smaller than for runs starting
with a flatter MF and the  trend between $\alpha$ and the
galactocentric distance is much weaker. As is evident from the data
in the table, there is a clear relationship between the amount of mass
which has escaped from the cluster and the variation of its mass function (we
will go through this point in greater detail later in the paper) and
the smaller changes in $\alpha$ for these runs simply reflect the
smaller mass loss rate for these systems.

We  now focus our attention on the runs with $\alpha_{i}=2.5,
W_0=7,M_i=6.15\times 10^4 \cmo$. Fig.1 shows the time evolution of
some global quantities for this set of runs. 
Panel (c), where the time
evolution of the total number of particles in  the first and the fifth bin
(corresponding to a range of masses $0.1<m/\mo<0.12$ and
$0.25< m/\mo<0.31$ respectively) of the mass function for
the run at $R_g=4 $Kpc and for that at $R_g=16$ Kpc has been
plotted, clearly  shows
the preferential escape of low-mass stars which is
responsible for the flattening of the IMF. As shown in panel (f) the
evolution of $\alpha$ is quicker for clusters at smaller
galactocentric distances and this gives rise to the correlation between
$\alpha$ and $R_g$. Finally it is interesting to note from panel (e)
that the fraction of the total mass of a system in white dwarfs
increases as the total mass of the cluster decreases (see sect.3.2 for
further discussion on all the issues concerning white dwarfs).

In fig.2 the MF at $t=0,7,10,15$ Gyr for main sequence stars for
the run at $R_g=4$ Kpc is plotted to show the progressive flattening of the
MF.

The data available and discussed above refer to a very limited set of
initial conditions; nevertheless, besides being important for the
indications they have provided, the results of these runs can be used
to derive some general analytical expressions for the main quantities
of interest for our work which will allow the results to be extended to a
larger set of initial conditions. From the data of these runs it is
possible to derive expressions providing the total mass of a cluster,
the slope of the MF and the fraction of white dwarfs at any time $t$, for any
value of the initial mass and any value of the galactocentric
distance within reasonable limits we will see below.
Let's start by investigating the behaviour of the total mass of the
cluster.
The total mass of a cluster decreases as a result of the mass loss due
to stellar evolution 
and two-body relaxation. For a given initial value of $\alpha$ (and
fixed values of the lower, $m_{low}$, and upper, $m_{up}$, limits of
the MF)  the fractional 
mass loss due to stellar evolution is the same for all clusters no
matter what their initial mass is or what  their galactocentric
distance is.
For the stellar evolution model adopted in this work this can be
explicitly written as a function of the stellar mass at
turn-off, $\tilde m$.
Having defined the following functions
\begin{eqnarray}
&& F_1(\tilde m)= {m_{up}^{2-\al}-\tilde
m^{2-\al}\over m_{up}^{2-\al}-m_{low}^{2-\al}}
-1.4\left({2-\al \over 1-\al}\right)\left({m_{up}^{1-\al}-\tilde
m^{1-\al}\over
m_{up}^{2-\al}-m_{low}^{2-\al}}\right)~~\hbox{for $8<\tilde
m/\mo<m_{up}$} \nonumber \\ 
&& \nonumber \\
&& F_2(\tilde m)={8^{2-\al}-\tilde m^{2-\al}\over 
m_{up}^{2-\al}-m_{low}^{2-\al}}~~\hbox{for $4.7<\tilde m/\mo<8$}  \\
&& \nonumber \\
&& F_3(\tilde m)=  0.78\left({4.7^{2-\al}-\tilde m^{2-\al}\over
m_{up}^{2-\al}-m_{low}^{2-\al} }\right)
-0.36\left({2-\al \over
1-\al}\right)\left({4.7^{1-\al} -\tilde m^{1-\al}\over
m_{up}^{2-\al}-m_{low}^{2-\al}}\right)  ~~\hbox{for $m_{low}<\tilde 
m/\mo<4.7$}\nonumber\\ 
\nonumber
\end{eqnarray}
we can write the fractional mass loss due to stellar evolution as

\begin{equation}
{\Delta M_{st.ev.} \over M_i}=\left\{
\begin{array}{ll}
F_1(\tilde m)& \mbox{if $8 \leq \tilde m/\mo<m_{up}$}\\
&\\
&\\
F_1(8)+F_2(\tilde m)& \mbox{if $4.7 \leq\tilde m/\mo< 8 $}\\
&\\
&\\
F_1(8)+F_2(4.7)+F_3(\tilde m)& \mbox{if $m_{low}\leq\tilde m/\mo<4.7 $}\\
\end{array}
\right.
\label{mlst}
\end{equation}

To calculate the total mass loss due to stellar evolution at any time
$t$ it is necessary to know the  turn-off mass at that time;
this is derived by interpolating the data provided in Table 1B of
Chernoff \& Weinberg (1990).
In fig.3a the time evolution of the total mass is presented. We plot
separately that due to mass loss associated with stellar evolution and
that due to other processes, i.e. mainly two-body relaxation and the
effect of the tide.

The latter shows clearly its
dependence on the initial conditions of the  mass loss due to
the latter process and its linear dependence on time. We point out that
while  the amount of  mass loss due to stellar evolution obtained by
eq. (\ref{mlst}) does not consider the possibility that some stars
might escape from the cluster before losing their mass, the curves
obtained from $N$-body data include only mass loss from stars inside
the tidal boundary of the cluster; for the IMF chosen in our
simulations the great majority of escaping stars have masses smaller
than those evolving before  20 Gyr and thus the theoretical estimate
and the data from simulations are almost coincident.

In fig.3b  we show that, scaling the time
by the parameter $F_{cw}$,
all the curves representing the time evolution of the
total mass (considering only the mass loss by two-body relaxation)
at different galactocentric distances coincide.
Thus we can write
\be
{M(t) \over M_i}=1-{\Delta M_{st.ev.} \over M_i}-{\beta\over F_{cw}}t
\label{mlo}\ee
where $\beta$ is a constant that can be determined by a linear fit of
the curves 
shown in fig.3b and is found to be $\beta\simeq 0.828$ (calculated as
the mean value of the slopes of the curves shown in figure 3b where
time is measured in Myr).
Eq.(\ref{mlo}) allows us to calculate the mass of a cluster at any
time $t$ for a 
cluster with initial mass $M_i$ and located at a distance from the
galactic center $R_g$; only a slight difference in the 
value of $\beta$ has been found for systems starting with $W_0=5$,
which in this case is $\beta\simeq 0.790$.
Fig.4 shows a 3-d plot of $M(15 Gyr)/M_i$ as a function of
$M_i$ and $R_g$ and a contour plot of this function.
The more massive clusters and those located at larger distances from
the galactic center are those which preserve a larger fraction of their
initial mass as two-body relaxation is less and less efficient and the
only mass loss is due to stellar evolution.

Though it is not the purpose of this paper to consider the lifetimes
of the globular clusters, it interesting in passing to compare our
result with some others which have been used in discussions of the
evolution of the Galactic globular cluster system. It is shown in
Aarseth \& Heggie (1997a) that the lifetime obtained by the methods we
use are in satisfactory agreement with those determined by Chernoff \&
Weinberg (1990) using the Fokker-Planck method. Aguilar et al. (1988)
used two different expressions for the lifetime (due to escape by
two-body relaxation) of a tidally bound cluster, but only for clusters
with stars of equal mass. We find that the lifetimes for a typical
model given by their formulae exceed our by a factor of about three or
four. Chernoff, Kochanek \& Shapiro (1986) also neglected stellar
evolution but did include several additional mechanisms such as disk
shocking. Even so their lifetimes exceed ours (which in this section
do {\sl not} include disk shocking) by a factor of two in typical
cases. Typical lifetimes to evaporation (by two-body effects alone,
but with stars of equal mass) given by Gnedin \& Ostriker (1997)
exceed ours by about 40 percent.

In order to obtain an analytical expression for the slope of the mass
function it is necessary to estimate the evaporation rate for stars
with different masses.
If the mass function is $dN \propto m^{-\al}dm$ we can write for any
values of the mass of stars $m_1$ and $m_2$
\be \alpha=- {\ln (dN/dm)_1-\ln(dN/dm)_2 \over \ln m_1-\ln m_2}.\ee
We estimate $(dN/dm)_{1,2}$
\be \left({dN\over dm}\right)_{1,2}={N_{1,2}(t)\over\Delta m_{1,2}}\ee
where $N_{1,2}(t)$ is the total number of stars with mass between
$m_{1,2}$ and $m_{1,2}+\Delta m_{1,2}$ at time $t$.
It is found from our results that the time evolution of total number
of particles with a given mass is 
approximately linear and the slope is inversely proportional to the
parameter $F_{cw}$
\be
N_1=N_1(0)\left(1-{\xi_1t\over F_{cw}}\right)
\ee
It follows that the slope of the mass
function at time $t$ can be written as
\be
\al(t)=\al(0)-{\left[\ln\left(1-{\xi_1 t\over 
F_{cw}}\right)-\ln\left(1-{\xi_2 t\over F_{cw}}
\right)\right]\over \ln m_1-\ln m_2} \label{nt}
\ee
We have chosen $m_1=0.1 \mo$ and $m_2=0.5\mo$ (this being the range where
our values of $\al$ are calculated).
 $\xi_1$ has been estimated by a linear fit of the
curve $N_1(t)$ for the run with $R_g=16$ Kpc (for $5~Gyr<t<10 ~Gyr$
where it has a definite linear behaviour). 
By fitting the values of $\al(15Gyr)$ from our runs at different
 $R_g$ with the function (\ref{nt}) we have  determined $\xi_2$.
Thus we get the following estimate for the slope of the mass function at
time $t$ for a cluster with initial mass $M_i$ 
orbiting at a distance $R_g$ from the galactic center 
\begin{eqnarray}
\al(t,R_g,M_i)=&&2.5+0.62\left[\ln\left(1-1.1506\left({t
\over Myr}\right)\left({ Kpc \over R_g}\right)\left({\cmo\over M_i
}\right)\ln (M_i/0.28)\right)  \right.\nonumber
\\ 
&&\left. \right. \nonumber \\
&&\left. \right. \nonumber \\
&&\left.
-\ln\left(1-0.2715\left({t
\over Myr}\right)
\left({ Kpc \over R_g}\right)\left({\cmo\over M_i}\right)\ln
(M_i/0.28)\right)\right]  \label{nemp} 
\end{eqnarray}
where we have taken $v_c=220 \hbox{km s}^{-1}$ and $0.28 \mo$ is the
initial mean stellar mass for the IMF adopted 
in our work. 
By eq.(\ref{mlo}) it is also possible to calculate  $\al$ at   any
given time $t$, 
 for a cluster located at $R_g$ and  whose mass at time $t$ is $M(t)$.
Fig.5 shows the curves of  $\al(15\hbox{Gyr})$ as a function of the
galactocentric 
distance for clusters having the same initial mass. Eq.(\ref{nemp}) is
valid as long as the number of particles is a linear function of time;
it is possible to see from fig. 1a  that towards the end of the cluster
lifetime, the behaviour of $N(t)$ deviates from the linear scaling with
time and slows significantly. The linear regime ends approximately at
$t \sim 8.5 \times 10^{-4} F_{cw}$ Gyr (or equivalently when $M_f/M_i
\sim 0.11$); for values of $t$ larger than this, eq.(\ref{nemp}) will
provide a slope of the MF smaller (i.e. a flatter MF) than the real
one.

Thus, in the plane $\alpha-R_g$, stellar mass loss and evaporation
due to two-body relaxation cause  points , initially on
the line $\al(R_g)=2.5$, to spread into the strip shown in fig.5.

While the curves on this diagram are qualitatively consistent with the
correlation between $\al$ and $R_g$ observed for Galactic globular
clusters (see below), much caution would be needed in the use of these
results in any {\sl detailed} interpretation of observations. In the
first place the observational data themselves are subject to
considerable uncertainty from a variety of sources, including the
distinction between global and local luminosity functions, the
questionable reliability of ground-based data, and our poor knowledge
of the mass-luminosity relationship for low-mass stars. Next, there
are limitations in our modelling, including the use of circular orbits
and a single initial mass function, and measurement of the mass
function over a single range in mass. Also, we have found that the
time scale on which the MF evolves is heavily determined by the
initial total mass, which is unknown for observed clusters.

One assumption of our models we have been able to check is the
influence of primordial binaries, by comparison with models discussed
by Aarseth \& Heggie (1997b in preparation). Inclusion of 10 percent
hard binaries by 
number appears to decrease $\alpha$ by an amount of order 0.05
(i.e. the mass function is flattened); this is small compared to the
evolutionary changes and fluctuations in the data (cf. Fig.6), and may
not be significant.

Despite the caveats it is still interesting to compare our
results with a representative sample of mass function slopes from the
literature. For example Capaccioli et al. (1993) have summarised
ground-based data for 17 clusters and they find a dependence of $\al$
on the current value of $R_g$ which is quite similar (qualitatively
and quantitatively) to that of Fig.5 for relatively low initial masses
($\ltorder 10^5 M_\odot$). To go further it would be necessary to give
attention to most caveats stated above. For example, Fig.13 (below)
shows that disk shocking increases the evolution of $\al$ beyond that
indicated in Fig.5, while our values of $\al$ are measured for stars
of lower mass than is the case with the observational
data. Furthermore, it might be more realistic to compare our
simulations with observational data for the dependence of $\al$ on the
estimated {\sl perigalactic} distance of the clusters.

Incidentally it is often implied that the observed correlation is a
signature of disk shocking. We have, however, shown that relaxation by
itself can significantly affect the shape of the IMF in a manner
qualitatively consistent with observation.

\subsection{Local mass function}
We turn now our attention to the differences between the
global/initial mass function and the local mass function measured at
various distances from the cluster center.
 Since in some  cases
observational data are taken for different clusters  at
different distances from the cluster center  it is important
to understand how large the difference between a local MF
and the global one can be  and where, inside the cluster, this difference is
larger, as well as to establish whether and where the PDMF keeps memory
of the IMF.

We have calculated the slope of the MF in three shells: the innermost
shell extends from the center to $r_{30}$, where  $ r_{30}$ is
the radius containing the innermost $30\%$ of the total number of
stars; the intermediate shell extends from $r_{30}$ to $r_{60}$ and
the outermost shell extends from $r_{60}$ to $r_{100}$ (both 3-d
and 2-d (projected distance) shells have been considered).
The limits of the three shells compared with the half-mass radius of
the cluster, $r_h$, are approximately  $(0-0.8r_h),(0.8-1.6r_h),(1.6-4r_h)$.
We focus our discussion on the results for the run at $R_g=4$ Kpc and for that at
$R_g=16$ Kpc. The former undergoes a strong mass loss and evolution of
the global value of $\al$ while, in the latter, mass loss due to two-body
relaxation is negligible and $\al$ does not change significantly.

If only mass segregation takes place and no star escapes from the
cluster due to relaxation, the mass function of the innermost shell
becomes flatter during the evolution  while the MF of the outermost one
becomes steeper. What happens when there is some mass loss
depends on how strong this is and on how efficient it is in
counteracting the trend given by mass segregation. In fact, while the
MF of the  inner shell will always tend to become flatter, the shape of
the MF of the outermost shell depends on the relative efficiency
of two processes acting in opposite directions: mass segregation
tends to steepen the MF, evaporation tends
to flatten it down. 
Figs. 6a-d show the time evolution of the MF in 2-d and 3-d shells  for
the runs at $R_g=4$ Kpc and $R_g=16$ Kpc.
The above qualitative scenario is evident from figs. 6a-b, relative
to the run at $R_g=4$ Kpc:
the MF of the innermost shell immediately flattens, that of the intermediate
shell  initially preserves its initial slope
but eventually flattens, and the MF of the outermost shell after an 
initial stage
during which tends to undergo a slight steepening eventually flattens
when mass loss effects dominate mass segregation. 

At $R_g=16 $ Kpc the only relevant effects are those due to mass
segregation and, as shown in figs. 6c-d, the slope of the MF
is flatter than the initial one in the inner shells and steeper than
the initial MF in the halo. We note that in this case no significant
mass loss occurs and the MF near the half-mass radius resembles quite
well the IMF (Richer et al. 1991, by an analysis of multi-mass
King models, arrived at the same conclusion), but  we
emphasize that the results for the run at $R_g=4$ Kpc, show this is
not always the case, since, if a significant mass loss takes place,
the PDMF at $r_h$ has no relationship with the IMF after a few Gyr.

To provide a quantitative estimate of the difference between the
initial/global MF and the local MF we have calculated the following
quantities
\be
{\Delta \al_{glob}}=\al_{shell}-\al_{glob},
\ee
and 
\be
{\Delta \al_{i}}=\al_{shell}-\al_{i}.
\ee
A negative (positive) value of these quantities means that the local MF is
flatter (steeper) than the global/initial MF.
The time evolution of the above quantities for the 3-d shells (
no qualitative difference is present in their behaviour for the 2-d
shells) is shown in figs 7a-d. At $R_g=4$ Kpc  a strong
mass loss takes place which eventually makes the MF of all the shells
flatter than the IMF. On the other hand when the local MFs are
compared with the global MF, the effects of mass segregation are
evident  as the outermost shell MF and the innermost 
shell MF are respectively steeper and flatter than the global one. The
intermediate shell MF quite well resembles the global one during the
entire evolution. At $R_g=16$ Kpc the mass loss due to two-body
relaxation is negligible and the global MF barely changes during the
entire simulation. In this case mass segregation is the only process
at work and the results obtained are in agreement with what is expected.

From the above results we can draw the following conclusions:
1) the MF in the outer shells of a cluster is always steeper than the
global one but it can be much flatter than the IMF if strong mass loss
occurred, thus changing it significantly; this means that observational data
taken in the outer regions of a cluster may be of no help in getting any
information about the IMF  if the cluster has
undergone a significant dynamical evolution; 2) the MF near the
half-mass radius is the most similar to the global one and thus it can
be significantly different from the MF in more external regions of the
cluster, always in the sense that the external MF is steeper; much
caution should be used in comparing observational data taken at
different distances  for different clusters and modelling of
mass segregation  effects (see e.g. King, Sosin \& Cool 1995) when any
comparison is to be done  is necessary.Though it is not the purpose of
this paper to investigate mass 
segregation as such, it is of interest to see how well its effect on
the local mass function slope can be modelled with a multi-mass
anisotropic King model. For the sake of illustration we took a model
with initial mass $1.49 \times 10^5 M_{\odot}$ and IMF index
$\alpha=2.5$ at galactocentric radius $R_g=4.4 $ kpc, and examined its
structure at 20 Gyr. This is a post-collapse model which has lost a
little over half its mass and about 20 percent of its remaining mass
is in the form of white dwarfs (see section 3.3). The stars were
binned uniformly in $\log M$ in the range $0.1M_\odot<m<M_\odot$ and
in Lagrangian radii corresponding to 10 uniformly spaced bins by total
number. The white dwarfs were added to the bin of stars with average
mass about $0.45 M_\odot$. Next a King model was selected to fit the
central parameters of all species and, by variation of the remaining
free parameters, the density profile of the stars of average mass
about $0.56 M_\odot$. Finally the spatial profile of $\alpha$ in the
King model, computed as in the $N$-body models, was compared with the
$N$-body result. It was found that the value of $\alpha$ in the King
model was slightly but sistematically smaller (by an average of about
0.4) than in the $N$-body model, except in the inner 10 percent, where
the models were forced to agree. Undoubtedly a post-collapse model is
a strenuous test of model fitting, and such a significant discrepancy
would not be expected in less evolved models.

Any difference in the MFs
of two clusters observed at different distances from the center does not
necessarily imply  their global MF are different and in any case 
their difference is likely not to signify the real
difference between the global MFs.
\subsection{ White dwarfs}
The code we have used for our investigation includes the effects of
stellar evolution according to the same model used in Chernoff \&
Weinberg (1990). Besides studying the effects of this process on the dynamical
evolution of a cluster it is thus possible to address the issue of the
presence of degenerate remnants in the cluster, white dwarfs and
neutron stars. 
As for the neutron stars these are found to be a very small fraction
of the total mass after 15 Gyr and no systematic investigation has
been possible due to the small numbers involved.
It  has been possible to do much more for the white dwarfs.
In table 1 the ratio of the total mass in white dwarfs to the total
mass at $t=15$ Gyr for all the runs done is shown. The first point
coming out from the data is that the smaller the
final value of the total mass, the larger  the fraction of the
mass in white dwarfs. The fraction of mass in white dwarfs results
from two opposite processes: the production of white dwarfs from stars
with initial mass less than 4.7 $\mo$ evolving from the main
sequence and the loss by evaporation through the tidal boundary of the
white dwarfs  which have already formed or, possibly, of their main sequence
progenitors.
For a power-law IMF with lower limit, $m_{low}$,
and upper limit $m_{up}$, the ratio of the
 total mass in white dwarfs, $M_{wd}$, produced at a
time $t$, when the turn-off mass is $\tilde m$, to the initial mass
$M_i$ is given by
\begin{equation}
\left({M_{wd}\over M_i}\right)_{prod}=\left\{
\begin{array}{ll}
0& \mbox{if $\tilde m/\mo>4.7$}\\
&\\
&\\
{2-\alpha \over m_{up}^{2-\alpha}-m_{low}^{2-\alpha}}
\left[{0.36 \over
1-\alpha}(4.7^{1-\alpha}-\tilde m^{1-\alpha})+{0.22 \over
2-\alpha}(4.7^{2-\alpha}-\tilde m^{2-\alpha})\right]& \mbox{if $\tilde
m/\mo<4.7$}. 
\end{array}
\right.
\label{wdt}
\end{equation}
The above expression would provide the total mass of white dwarfs if no
loss of stars took  place. Actually a certain
fraction of  the white dwarfs produced  escapes and this expression 
provides only an upper limit for $M_{wd}$.
The real value of $M_{wd}/M_i$ can be written as the product of the
eq.(\ref{wdt}) by a function of time taking into account the reduction
of the expected mass in white dwarfs due to the loss of stars through
the tidal boundary
\be
\left({M_{wd}\over M_i}\right)_{real}=\left({M_{wd}\over
M_i}\right)_{prod}\times F(t)
\label{wdr}
\ee
The function $F(t)$ has to be determined by the results of $N$-body
simulations. 
In fig. 8a we show the time evolution of $M_{wd}/M_i$ for the runs
starting with $W_0=7$, $\al_{i}=2.5$ and $R_g=4,5,8,16$ Kpc. Also
shown in the 
figure is $\left({M_{wd}\over M_i}\right)_{prod} $ which, as expected,
always lies above the curves corresponding to the data from $N$-body
simulations. 
The function $F(t)$ obtained from the ratio of  $M_{wd}/M_i$ from
$N$-body data to $\left(M_{wd}/ M_i\right)_{prod} $ (eq.\ref{wdt})
is shown for the same runs  in fig. 8b.
In fig. 8c the same function, this time for all the runs (both
$W_0=5$ and $W_0=7$), versus the time scaled by the
parameter $F_{cw}$ is shown. As expected some spread is still  present
among curves 
corresponding to different initial conditions; in fact even though the
effect of two-body relaxation in causing stars to evaporate from a
cluster is the same for the same value of the scaled time, the properties
of the population of white dwarfs (their mass distribution
essentially) is different and a difference in the fraction of escaped
white dwarfs is thus expected. Nevertheless this difference is not very large
and we can approximately write $F(t)$ as a unique function of the scaled
time, $t/F_{cw}$.
In order to obtain an analytic expression for this function, we have done a
polynomial fit of the data  obtained from  the two runs at $R_g=4$ Kpc
starting with $W_0=7$ and $W_0=5$ (fig.8d)
\be
F(t,R_g,M_i)=F({t/F_{cw}})=1-0.794{t\over F_{cw}}+7.12 \times
10^{-5}\left({t\over F_{cw}}\right)^2
-3.82 \times
10^{-9}\left({t\over F_{cw}}\right)^3
\label{polfit}
\ee
which provides a good approximation for
$t<9\times 10^{-4} F_{cw}$ 
Gyr (or equivalently for $M_{fin}/M_i>0.07$).
By the above equations we can predict the fraction of the total
(initial or at any time $t$) mass 
in white dwarfs  
at any time $t$ for clusters at a galactocentric
distance $R_g$ (Kpc) with an initial mass $M_i~(\cmo)$ or equivalently
for clusters 
having a family parameter $F_{cw}=M_iR_g/\ln(M_i/0.28)$ (0.28 being
the mean value of the mass for the IMF we are considering and for
which the analytic expression of $F(t,R_g,M_i)$ is valid).

In fig. 9 we show the plot of $M_{wd}/M(15 Gyr)$ versus
 $F_{cw}$:  $M_{wd}/M(15 Gyr)$ is a decreasing function of
$F_{cw}$ and this means that the fraction of the total mass in white dwarfs is
larger for clusters undergoing a strong dynamical evolution and losing
a large fraction of their initial mass;
the smaller the ratio of the final to the initial total  mass is, the
larger the 
fraction of the final mass in white dwarfs is. Fig.10 shows 
the  plot of $M_{wd}/M(15 Gyr)$   versus the logarithm
of the total mass of the cluster at $t=15$ Gyr for different values of
$R_g$.
Not very much is currently known about the fraction of white dwarfs in
globular cluster from observations; the only estimates are supplied by
authors using observed light (and velocity dispersion) profiles to
model the
structure and the stellar content of individual globular clusters
usually by multi-mass King models ( see e.g. Meylan 1987 and 
references therein).
 
Following this method, the fraction of white dwarfs is estimated
assuming that the 
current value of the slope of the MF is equal to that of the IMF and,
consistently, that the only mass loss that occurred during the cluster
lifetime is due to stellar evolution (hereafter we will refer to this
estimate as the 'observational' estimate). It is clear that if mass loss 
due to  relaxation takes place and, as a consequence  of this
 the IMF is different from
the present one, this procedure will not provide a correct estimate of the
content of white dwarfs. Assuming that the slope of the IMF is equal
to $2.5$, we can use the results of our simulations to provide a
quantitative estimate of the error incurred.
Fig.11a shows the contour plot of $M_{wd}/M(t)$ in the plane
$t-\al(t)$ providing the value of $M_{wd}/M(t)$ for any cluster once
the slope of its mass function and the time this value of the slope is
reached are known (assuming the initial value of $\al$ is equal to
2.5); dashed lines show the evolution of $\al$ for clusters with
different values of $F_{cw}$ obtained from eq. (\ref{nemp}).
Fig.11b shows an analogous plot but in this case the values of
$M_{wd}/M(t)$ are the 'observational' estimates.
Finally fig.11c shows the contour plot of the ratio of the two above
estimates of  $M_{wd}/M(t)$ (the estimate from $N$-body data to the 
observational one) and it provides  the
correction to be made to the ' observational' estimate; 
the latter overestimates  the content of white dwarfs but
the error made for the fraction of white dwarfs at $t=15$ Gyr is never
too large, as the correction to be made is never smaller than about
0.65 at $t=15$ Gyr.
Figs. 12a-c show plots analogous to those of figs. 11a-c but in the
plane $\log M(15 \hbox{Gyr})$-$R_g$.
\section{Results:disk shocking}
In this section we will show the results of a set of simulations
including  the effects of disk shocking.
The main goal is that of  investigating to what extent this process can
alter the results described in the previous section in which
 only stellar evolution and  relaxation were considered.

We begin by making some preliminary qualitative considerations about
 the expected changes due to disk shocking
 in the three quantities (total mass, white
dwarf content and MF slope) we have focussed our attention on until
now. 

As for the total mass, it is obvious that disk shocking  will cause
an additional mass loss  and  the total mass of a cluster at any time
$t$ will  be  (depending on the strength of disk
shocking) less than or equal to the corresponding value for a cluster
starting with the 
same initial conditions but not undergoing disk shocking.
It is not so obvious  what to expect  for the content of white
dwarfs; no simple qualitative argument can predict if the
same correlation between the fraction of total mass in white dwarfs and the
ratio of total to initial mass, shown to hold in the previous section,
still holds in this case. In fact, it is not possible to know, {\it
a priori}, if
the mass lost by disk shocking will contain the same  fraction
of white dwarfs it would contain if this additional mass loss were due to
two-body relaxation. Analogous information is required to predict
the effects of disk shocking on the evolution of the MF. As for this
latter issue one  difference between the mass
loss due to disk shocking and that due to two-body relaxation is that,
contrary to what happens for mass loss due to two-body
relaxation,
 disk shocking does not produce any differential escape of
stars with  different mass unless the masses are segregated by some
other mechanism; the change in the velocity of a star due to
 disk shocking, for a cluster at a given
galactocentric distance, depends only on the distance from the cluster
center and not on the mass of the star. This means that  only with the joint 
effect of mass segregation, driving low mass stars into  the halo and
high-mass stars into the core, can the mass loss due to disk shocking
alter the slope of the mass function. The extent to which disk shocking can
alter the mass function thus depends not only on the mass loss but
also on  mass segregation. 

Initial conditions 
and the  main results for the runs including disk shocking
are shown in table 2. 
Besides running simulations with the same initial
conditions adopted for runs without disk shocking,
 we have investigated
a set of initial conditions having the same galactocentric distance
but different values of the initial mass (and thus different values of
$F_{cw}$) and a set of initial conditions all having the same value of
$F_{cw}$ but different galactocentric distances. The former set have
been investigated in order  to determine the differences in the
effects of a sequence of interactions with a disk
 having the same strength (fixed $R_g$) on
clusters having different values of $F_{cw}$,  while the latter have
been done to
study the effects of disk shocking with a
varying strength on clusters all having the same initial value of
$F_{cw}$. The results of these runs, besides providing useful indications
on the effects of disk shocking on the stellar content of globular
clusters, should make possible  the derivation of more general expressions for
the total mass, the slope of the MF and the fraction of the total mass
in white dwarfs 
(eqs.(\ref{mlo},\ref{nemp},\ref{wdr})).

In fig.13 we have plotted  the slope of the MF after 15 Gyr as a
function of the 
galactocentric distance, showing to what extent disk shocking has
changed the final result expected for runs 
starting with the same initial conditions but not undergoing disk
shockings.

Besides the same quantities already shown in table 1 for the runs
without disk shocking, the fraction of the initial mass lost by disk
shocking $\Delta M_{ds}/M_i$ and the additional change in the slope of
the MF due to disk shocking have been included in table 2. Both of these
quantities have been calculated as the difference between the values
obtained from the simulations with disk shocking and the results
expected for the same initial conditions  without the effects of disk
shockings, the latter being calculated by
eqs.(\ref{mlo},\ref{nemp}).
In figs. 14a-b the evolution of the total mass and of the slope of the
MF for  the runs not including disk shocking is compared with that
obtained from runs with the same initial conditions and including the
effects of disk shocking. In agreement with what is expected, the
evolution of both  the total mass and the MF is faster for systems
undergoing disk shocking (except for the system at $R_g=16$ Kpc where
the effects of disk shocking are negligible and the differences are
due to statistical fluctuations).

These plots provide  generic information on the effects of disk
shocking, but the most important information to answer the questions
raised above 
concerning the fraction of white dwarfs contained in the mass lost by
disk shocking and the effect of this process on the evolution of the
slope of the mass function comes from the plots shown in figs. 15-16.

In these figures the fraction of the total  mass at $t=15 $ Gyr in
white dwarfs and the difference between the initial slope of the MF
and that  at $t=15 $ Gyr are plotted against the ratio of the
total mass at $t=15$ Gyr to the initial one; the values predicted from
the  analytical expression derived in the previous section for the
runs without disk shocking are also shown. Both the change in the
slope of the MF and the content of white dwarfs depend only on the
fraction of mass lost during the evolution no matter whether 
relaxation is the only process causing the evaporation of stars or
disk shocking is also responsible for a part of the escaping stars.
This means that the content of the fraction  of mass  lost by disk
shocking is similar to that lost by two-body relaxation.
We note that, even though the values of the final mass we have plotted
in fig.16 are determined  both by mass loss due stellar evolution and
relaxation/disk shocking/tidal stripping, obviously the relevant quantity in
determining the variation in the slope of the IMF (for main sequence
stars) is only the escape of stars due to relaxation/disk
shocking/tidal stripping.
Since the data plotted refer to systems all with the same IMF and thus
all losing by stellar evolution  the same fraction of their initial
mass,the difference between the real final mass and the final mass
calculated considering only the escape of stars by relaxation/disk
shocking is  equal to a constant ($\sim 0.18$ for the IMF adopted
in our work; in fact $\alpha_i-\alpha(15)$ tends to 0 for
$M_f/M_i\simeq 0.82$ that is when the only mass loss is due to stellar
evolution). Should data from systems losing a different fraction of
their initial mass by stellar evolution be considered for this plot,
this  correction should be properly taken into account.

The same remark applies to fig.15; in this case also the dependence
on the IMF of the total mass of white dwarfs produced should be considered.

If, analogously to what was  done for the runs without disk shocking, we could
get an analytical  expression for the time evolution of the mass,
 we might then use the relationship
between this quantity and the fraction of white dwarfs and the slope
of the MF to obtain these quantities at any time $t$ and for a quite
general set of initial conditions.

Fig. 17 shows the time evolution of the total mass for four different runs
 including disk shocking, two for initial conditions with the same
value of $F_{cw}$  and two for initial conditions having the same
galactocentric distances. The mass lost by stellar evolution has
been added so that the plotted lines show only the mass lost by disk
shocking and two-body relaxation. Similarly to what happens when disk
shocking is not included, the mass is a linear function of time.
This and the results coming from figs. 15-16 suggest that it is possible
for any initial condition to define an {\it equivalent family parameter},
$F_{cw}^{eq}$ having the following meaning:
the evolution of  a cluster with a family parameter $F_{cw}$ whose
evolution is driven by both two-body relaxation and disk shocking is
equivalent to that of cluster whose evolution is driven by two-body
relaxation only but  whose family parameter is $F_{cw}^{eq}$, where
$F_{cw}^{eq} \leq F_{cw}$ always. Of course $F_{cw}^{eq}$ depends on
$F_{cw}$ and on $R_g$ and the main goal is now that of finding out this
dependence by which we will be able to calculate analytically
 any quantity we are interested in once the initial conditions are
specified, exactly as we have done in the previous section.

In the previous section we have shown that the mass loss is a linear
function of time and that the mass loss rate is inversely proportional
to the family parameter $F_{cw}$; in order to get an analytic
expression for $F_{cw}^{eq}$ we have calculated from the results of
our $N$-body simulations an empirical
analytical expression for the mass loss rate as a function of $F_{cw}$
and $R_g$ and derived from this $F_{cw}^{eq}$.
Analogously to what was done for the runs without disk shocking, we can
write
\be
{M(t) \over M_i}=1-{\Delta M_{st.ev.} \over M_i}-\lambda t
\label{mlods};
\ee
as shown in the previous section, when two-body relaxation is the only
process considered, $\lambda$ depends only
on $F_{cw}$ ($\lambda=\beta/F_{cw}$, see eq.(\ref{mlo})); if the
effects of disk shocking also are taken into account $\lambda$ depends
both on $F_{cw}$ and on $R_g$. From our data we find
\be
\log \lambda=0.6931-1.46 \log R_g-1.134\log F_{cw}+0.2916\log F_{cw}
\log R_g \label{slp}.
\ee
The above expression is completely empirical and it has been derived
from data spanning
a limited range of values of $R_g$ ($1.1<R_g<16$) and
$F_{cw}$($2\times 10^4<F_{cw}<1.13\times 10^5$) but it nevertheless
provides useful qualitative and quantitative information on the
evolution of clusters including the effects of disk shocking.
The plot of $\lambda$ obtained from $N$-body data
versus the value calculated from eq.(\ref{slp}), fig. 18, shows that
eq.(\ref{slp}) approximates well the dependence of $\lambda$
on $F_{cw}$ and $R_g$.

As explained above, once $\lambda$ is known, we can easily calculate
$F_{cw}^{eq}$; fig.19a shows the ratio of the total mass at $t=15 $Gyr
to the initial mass versus $F_{cw}$ for all the runs
done, both with and without disk shocking, with the solid line showing
the values predicted from eq.(\ref{mlo}). As expected the real value of
$F_{cw}$ does not provide a good indication of the mass loss for the
runs with disk shocking; if $F_{cw}$ is replaced by $F_{cw}^{eq}$ (of
course no difference exists between these two quantities for runs
without disk shocking) all the data are located, with a
good approximation, along the curve predicted by eq.(\ref{mlo}) (fig.19b).

In fig. 20 families of models having given
values of $F_{cw}$ and $F_{cw}^{eq}$ are shown; the qualitative
behaviour agrees with what one would expect: 
two clusters located at the same galactocentric distance, one
undergoing disk shocking and one not, 
evolve in the same way if
 the initial mass of the former is larger than that of
the latter, 
while if they have the same initial mass, they  evolve in the same way
if  the former
is located at a galactocentric distance larger than the latter.

Analogous information is contained in fig. 21 where curves of
equal-mass at $t=15$ Gyr are shown for clusters evolving with and
without disk shocking in the plane $\log M_i-R_g$.

We can now calculate analytically $\alpha$ and the content of white
dwarfs at any time $t$ and for any initial condition (within the
limits mentioned in this and in the previous section and for the
particular choice of IMF adopted in our work) taking into
account also the effects of disk shocking. 

Fig. 22  shows the changes due to disk shocking
 in the curves of $\alpha(15 Gyr)$ versus
$R_g$ for different values of the initial mass.
As for the formation of a correlation between the slope of the mass function
and the galactocentric distance, as could be seen already
from fig. 13 , disk shocking has the effect of increasing the trend
established by two-body relaxation.
\section{Dependence of the results on $N$}
By the simulations described in the previous sections, starting with
$N=4096$ particles, we are 
trying to model the evolution of stellar systems typically having
$N\simeq 10^6$ stars; the time scaling adopted to convert time in
$N$-body units to astrophysical units should allow a proper use of the
$N$-body data to investigate the evolution of real clusters.
Nevertheless, as shown also in Aarseth \& Heggie (1997a), a slight dependence
of the results on $N$ still exists. In order to give a quantitative
estimate of the extent of the $N$-dependence
 we have done seven runs starting with the same initial
conditions $(W_0=5,R_g=4~\hbox{Kpc},M_i=6.15\times 10^4 \cmo)$ but with a
different initial number of particles (four runs with $N=4096$, two
runs with $N=8192$ and one run with $N=16384$ so as to have quantities
all with the same statistical significance for all the values of $N$
investigated).
In table 3 we summarize the relevant information on these runs.

In agreement with the results of Aarseth \& Heggie (1997a), we find
that the larger the number of particles is, the faster the
evolution of the system is. 
Part of the explanation of this  is likely to reside in the
differences in the 
structure of clusters with different $N$ at the end of the initial
phase dominated by mass loss due to stellar evolution. In fact, as
explained in larger detail in Aarseth \& Heggie (1997a), scaling
time  from $N$-body units to astrophysical units by the ratio of
$N$-body to real relaxation time implies that the smaller the number
of stars is, the smaller the ratio of time scale of mass loss due
stellar evolution to the crossing time scale is. This is quite
important since it is well known  (see e.g. Hills 1980, Lada, Marculis
\& Dearborn
1984, Fukushige \& Heggie 
1995, Aarseth \& Heggie 1997a) that the rate at which
mass loss occurs plays  an important role in
determining the evolution of a stellar system, spanning from an
expansion of the entire cluster proportional to $1/M$ without the
escape of any star in the case of
slow mass loss to the complete disruption of the cluster in the case
of rapid mass loss exceeding 1/2 of the entire initial mass of the
cluster; this difference is likely to produce a difference in the final
structure of clusters losing mass by stellar evolution in the
impulsive or adiabatic regime.  The subsequent differences in the mass loss
rate are likely to be due, at least partially, to this difference even
though further investigation on this point is needed.
Some simulations with $N=4096$ and $N=8192$ not including
stellar evolution have been carried out to test to what extent the
above effect was actually able to explain the observed differences.
The trend for system with larger values of $N$ to evolve faster has
been found to be still present, even though to a smaller extent.
This means that part of the differences observed are unlikely to be
due to the chosen scaling and their origin must be due to a real
difference in the evolution of systems with different $N$ not simply
scaling with the relaxation time. We could not find a convincing
explanation for this result but possibly the dependence on $N$ of the
depth of the potential well and the consequent dependence of the
fraction of stars ejected from the core reaching the outer regions
(see e.g. Giersz \& Heggie 1994) might explain the observed
differences.

Fig. 23, in which  we have  plotted the slope of $M(t/F_{cw})/M_i$,
i.e.  $\beta$, (excluding the mass loss due to stellar evolution)
versus $N$, clearly shows the increase in the mass loss rate as $N$
increases.
A correction of approximately  $20 \%$ seems to be necessary to the
conversion from time in $N$-body units to time in astrophysical units
adopted in the set of runs with $N=4096$.

As shown in figs. 15 and 16, however, the runs for all values of N
yield the same result for the way in which both the variation in the
slope of the mass function and the content of white dwarfs at $t=15$
Gyr depend on the fraction of initial mass left after 15 Gyr.

While further investigation on the $N$ dependence of the results is still
necessary, particularly for what concerns the origin of this
dependence, in the light of this last result it is clear that the
correction to be made just implies that our results from  simulations
with $N=4096$ at $t=15$ Gyr would
actually be more relevant for real clusters at $t=12$ Gyr and that results
from simulations at $t=18$ Gyr would be  those relevant for real
clusters at $t=15$ Gyr.

\section{Summary and conclusions}
In this work  we have carried out a large set of $N$-body simulations to
investigate the dynamical evolution of globular clusters, 
focusing our attention on the effects of
dynamical evolution on the stellar content of the systems; in
particular we have investigated the evolution of the MF and of the
fraction of the total mass in white dwarfs.
The code used includes the
effects of stellar evolution, two-body relaxation and disk shocking and
takes the presence of the tidal field of the Galaxy into account.
A set of different initial conditions for the structure of the cluster
and for the galactocentric distance has been considered.

The dependence of the slope of the MF and of the fraction of white
dwarfs on the initial conditions of the
clusters has been explored and in particular we have 
determined to what extent the observed correlation between the slope of
the MF and the galactocentric distance can result from the effects
of dynamical evolution; we have tried to provide a quantitative estimate of
the role played by relaxation and disk shocking.
The dependence of the slope of the mass function on the location
inside the cluster has been also investigated, our attention being focused
on the differences between the local MF, the IMF and the global
PDMF.

The main conclusions we can draw are:
\begin{enumerate}
\item in agreement with what is expected, as a result of  mass loss
through the tidal boundary, both 
due to two-body relaxation and to disk shocking, the global mass
function becomes flatter. For given initial parameters, mass loss is
stronger for clusters closer to the galactic center, and,
consequently, a trend between the slope of the MF and the
galactocentric distance forms as evolution goes on.
This trend is stronger for low-mass clusters, as these have shorter
relaxation times and thus evolve more quickly than massive clusters.
Both mass loss by two-body relaxation and disk shocking are important
in causing the 
MF to flatten.
By the results of $N$-body simulations we have derived
an analytical expression for the slope of the mass function at any
time $t$ and for any initial value of the mass and of the
galactocentric distance both with and without 
the effects of disk shocking.

The difference between the initial and the final (at $t=15$ Gyr) slope
of the MF has been shown to depend approximately only on the fraction
of the initial 
mass lost and this dependence is the same
no matter whether disk shocking is included or not.
\item The MF near the
half-mass radius is the one which, during the entire evolution, is least
affected by mass segregation and quite well resembles the present-day
global mass function. The extent to which the MF observed near the
half-mass radius 
can provide us with useful information on the IMF thus depends on the
difference between the IMF and the PDMF.
Possibly, when the the PDMF is  different from
the IMF, observations of the MF at radii larger than the half-mass radius
can supply indications on the IMF, but it is
important to note that, in cases of strong mass loss and evolution of
the IMF, we have
 shown that, even the MF in the outer regions of the clusters eventually
becomes flatter than the IMF whose memory is thus completely erased
from the PDMF.
The MF in the inner regions is always significantly flatter than the PDMF
as a result of mass segregation.
\item The ratio of the total mass of white dwarfs retained in the
cluster to the total mass of the cluster, $M_{wd}/M(t)$,
increases during the evolution: as the fraction of the
initial mass left in the cluster decreases, the fraction of this in
white dwarfs increases. 
The total mass of white dwarfs is determined by the interplay between
the rate of production determined by the time scales of stellar
evolution and the rate of escape through the tidal boundary
 of the  white dwarfs, or possibly of their
progenitors, determined by the relaxation time and disk shocking time
scale.
We have obtained an analytical expression for the fraction of white
dwarfs present in a cluster as a function of the initial conditions
and of time as the product of the production rate, that can be easily
derived analytically once the IMF has been given, by the escape rate,
 which is derived instead  from a fit to the $N$-body data.
Analogously to what was shown for the difference between the initial and
the final slope of the MF, the dependence of
 $M_{wd}/M(t)$ on $M(t)/M_i$ is the same no matter
whether the effects of disk shocking are included or not.

We have made a comparison between our estimate of the fraction of
white dwarfs and 
the one which would be  obtained by extrapolating the present day main
sequence mass function (a procedure often used in literature) and 
 we have shown that the former is, in most
cases, smaller than the latter, with the ratio of the two
estimates 
ranging, in most relevant cases, from about 0.65 to 1 depending on the
initial conditions. 
\item All the simulations we have carried out for our work started
with an initial number of particles $N=4096$. The dependence of our
results on $N$ has been checked by an additional set of simulations
starting with larger number of particles ($N=8192$, $N=16384$). In
agreement with the results by 
Aarseth \& Heggie (1997a) we have shown that the scaling of time from
$N$-body units to  astrophysical units adopted for $N=4096$ requires a
correction of about $20 \%$ when the results are to be used to model
systems with values of
 $N$ relevant for real globular clusters.

\end{enumerate}
\section*{Acknowledgments}
We thank Sverre Aarseth for the use of NBODY4 and for especially
tailoring it to the needs of this project.\\
We are most grateful to the referee, whose comments have helped to
improve the paper.\\
This work has been supported by SERC/PPARC under grant number
GR/J79461, which funded the GRAPE-4 (HARP) hardware.\\
EV acknowledges financial support by a ESEP fellowship from The Royal
Society and  Accademia Nazionale dei Lincei and the hospitality
of the Department of Mathematics and Statistics of the University of Edinburgh.

\section*{References}
Aarseth S.J., 1985, in Brackbill J.U., Cohen B.I. eds., Multiple Time
Scales. Academic Press, New York, p.377\\
Aarseth S.J., 1994, in Contopoulos G., Spyrou N.K., Vlahos L. eds.,
Galactic Dynamics and N-body Simulations, Springer Verlag, p.277\\
Aarseth, S.J., Heggie, D.C., 1997a, submitted to MNRAS\\
Aarseth, S.J., Heggie, D.C., 1997b, in preparation\\
Aguilar L., Hut P., Ostriker J.P., 1988, ApJ, 335, 720\\
Alexander, D.R.,Brocato E., Cassisi S., Castellani V., Ciacio F.,
Degl'Innocenti S., 1997, A\&A, 317,90\\
Bahcall J.N., 1984, ApJ, 287, 926\\  
Bahcall J.N., Schmidt M., Soneira R.M., 1982, ApJL, 258, L23\\
Capaccioli, M., Ortolani, S., Piotto, G., 1991, A\&A, 244,298\\
Capaccioli, M.,Piotto, G., Stiavelli, M., 1993, MNRAS, 261, 819\\
Chandrasekhar, S., 1942, Principles of Stellar Dynamics. Univ. of
Chicago Press, Chicago\\
Chernoff, D.F., Kochanek, C.S., Shapiro, S.L., 1986, ApJ, 309, 183\\
Chernoff, D.F., Weinberg, M., 1990, ApJ, 351, 121\\
D'Antona, F., Mazzitelli, I., 1996, ApJ, 456, 329\\
Dauphole, B., Geffert, M., Colin, J., Odenkirchen, M., Tucholke, H.J.,
1996, A\&A, 313, 119\\ 
Djorgovski, S., Piotto,G., Capaccioli, M., 1993, AJ, 105, 2148\\
De Marchi, G., Paresce, F., 1995a, A\&A, 304,202\\
De Marchi, G., Paresce, F., 1995b, A\&A, 304,211\\
Drukier G., 1996, MNRAS, 280, 498\\
Elson, R., Gilmore, G., Santiago, B., Casertano, S., 1995, AJ, 110,682\\
Fukushige, T., Heggie, D.C., 1995, MNRAS, 276, 206\\
Gebhardt K., Pryor C., Williams T.B., Hesser J.E., Stetson P.B., 1997,
AJ, 113, 1026\\
Giersz, M., Heggie, D.C., 1994, MNRAS, 270, 298\\
Giersz, M., Heggie, D.C., 1997, MNRAS, in press\\
Gnedin O., Ostriker J.P., 1997, ApJ, 474, 223\\
Heggie,D.C., Mathieu, R.D., 1986, in Hut P., McMillan, S.L.W. eds.,
The Use of Supercomputers in Stellar Dynamics. Springer-Verlag,
Berlin, p.233\\
Heggie D.C., Hut P., 1996,in Dynamical Evolution of
Star Clusters, IAU Symp. 174, eds. P.Hut and J.Makino, p.303\\ 
Hills J.G., 1980, ApJ, 235, 986\\
Iben I.J., Renzini A., 1983, ARA\&A, 21, 271\\
King, I.R., 1966, AJ, 71, 64\\
King, I.R., 1996, in Dynamical Evolution of
Star Clusters, IAU Symp. 174, eds. P.Hut and J.Makino, p.29\\
King, I.R., Sosin, C., Cool,A.M., 1995, ApJL, 452, L33\\
Lada C.J., Margulis M., Dearborn D.S., 1984, ApJ, 285, 141\\
Makino, J., Kokubo, E., Taiji, M., 1993, PASJ, 45, 349\\
McClure, R.D.,et al., 1986, ApJ, 307,L49\\
Meylan, G., 1987, A\&A, 184, 144\\
Paresce, F., De Marchi, G., Romaniello, M., 1995, ApJ, 440, 216\\
Piotto, G., 1991,in Formation and Evolution of Star Clusters, ed.
K.Janes, ASPCS vol.13, p.200\\ 
Piotto, G., Cool, A.M., King I.R., 1996, in Dynamical Evolution of
Star Clusters, IAU Symp. 174, eds. P.Hut and J.Makino, p.71\\
Piotto, G., Cool, A.M., King I.R., 1997, AJ, in press\\
Richer, H.B., Fahlman, G.G., Buonanno, R., Fusi Pecci, F., Searle, L.,
Thompson, I.B., 1991, ApJ, 381,147\\
Richer H.B., Fahlman G.G., Ibata R.A., Stetson P.B., Bell R.A., Bolte
M., Bond H.E., Harris W.E., Hesser J.E., Mandushev G., Pryor C.,
Vandenberg D.A., 1995, ApJ, 451, L17\\
Richer H.B., Fahlman G.G., Ibata R.A.,Pryor C.,Bell R.A., Bolte M.,
Bond H.E., Harris W.E., Hesser J.E., Holland S., Ivanans N., Mandushev
G., Stetson P.B., Wood M.A., 1997, ApJ, submitted\\
Santiago, B.,Elson, R., Gilmore, G., 1996, MNRAS, 281, 1363\\
Spitzer, L., 1958, ApJ, 127, 17\\
Spitzer, L., 1987, Dynamical evolution of globular clusters, Princeton
University Press\\
Stiavelli, M., Piotto, G.,Capaccioli, M., Ortolani, S., 1991,in
Formation and Evolution of Star Clusters, ed. 
K.Janes, ASPCS vol.13, 449\\
Stiavelli, M., Piotto, G.,Capaccioli, M., 1992, in Morphological and
Physical Classification of Galaxies, ed. G.Longo, M.Capaccioli and G.
Busarello, p.445\\
Weinberg, M., 1994, AJ, 108, 1414\\
\newpage
\clearpage
\begin{table}
\begin{tabular}{|ccccccccc|}
\multicolumn{9}{c}{\bf Table 1}\\
\multicolumn{9}{c}{Results of the simulations including two-body
relaxation and 
stellar evolution}\\
\hline
$M_i(M_{\odot})$&$ R_g$ (Kpc)&$F_{cw}/10^4$&$ W_0$& $\alpha_{i}$&$
M_f/M_i$&$ M_{wd}/M_f$&$ \alpha_f$&$ \alpha_i-\alpha_f$\\
\hline
$6.15\times 10^4$& 4&$ 2.0 $&7& 2.5& 0.205& 0.279& 1.46& 1.04\\
$6.15\times 10^4$& 5&$ 2.5$& 7& 2.5& 0.307& 0.199& 1.81& 0.69\\
$6.15\times 10^4$& 8&$ 4.0$& 7& 2.5& 0.524& 0.149& 2.21& 0.29\\
$6.15\times 10^4$& 16&$ 8.0$& 7& 2.5& 0.657& 0.128& 2.41& 0.09\\
\hline
$6.15\times 10^4$& 4&$ 2.0 $& 5& 2.5& 0.195& 0.273& 1.38& 1.12\\
$6.15\times 10^4$& 5& $2.5 $& 5& 2.5& 0.344& 0.196& 1.96& 0.54\\
$6.15\times 10^4$& 8& $4.0 $& 5& 2.5& 0.542& 0.150& 2.31& 0.19\\
$6.15\times 10^4$& 16& $8.0$& 5& 2.5& 0.670& 0.130& 2.39& 0.11\\
\hline
$6.15\times 10^4$& 4 & $1.92$& 7& 3.5& 0.633& 0.013& 3.27& 0.23\\
$6.15\times 10^4$& 5& $2.4 $& 7& 3.5& 0.702& 0.012& 3.39& 0.11\\
$6.15\times 10^4$& 8&$ 3.84$& 7& 3.5& 0.808& 0.014& 3.42& 0.08\\
$6.15\times 10^4$& 16&$ 7.68$& 7& 3.5& 0.890& 0.019& 3.46& 0.04\\
\hline
\multicolumn{9}{l}{Final values are calculated at $t=15$ Gyr}\\
\multicolumn{9}{l}{$M_{wd}$ is the total mass of white dwarfs in the
cluster at $t=15$ Gyr.}\\
\end{tabular}
\end{table}
\newpage
\clearpage
\begin{table}
\begin{tabular}{|ccccccccc|}
\multicolumn{9}{c}{\bf Table 2}\\
\multicolumn{9}{c}{Results of simulations including two-body
relaxation, stellar evolution and disk shocking}\\
\hline
$M_i(M_{\odot})$&$R_g$ (Kpc)&$F_{cw}/10^4$&$ M_f/M_i$&$ M_{wd}/M_f$&$
\alpha_f$&$ \alpha_i-\alpha_f$& $\Delta \alpha_{ds}$& $\Delta M_{ds}/M_i$\\
\hline
$6.15\times 10^4$& 4&$ 2.0$& 0.080& 0.340& 0.37&
2.13&1.05& 0.113\\ 
$6.15\times 10^4$& 5&$ 2.5$& 0.254& 0.242& 1.50&
1.00&0.39& 0.064\\ 
$6.15\times 10^4$& 8& $4.0$&0.489& 0.149& 2.18& 0.32&
0.04& 0.015\\ 
$6.15\times 10^4$ &16&$ 8.0$& 0.670& 0.127& 2.38& 0.12&
0.002& -0.011\\ 
$1.59\times 10^5$& 2.1&$ 2.5$& 0.166& 0.292& 1.12& 1.38&
0.77& 0.156\\  
$1.59\times 10^5$& 3.3& $4.0$&0.418& 0.168& 2.10& 0.40&
 0.11& 0.083\\ 
$1.59\times 10^5$& 6.7&$ 8.0$& 0.610& 0.123& 2.35& 0.15&
 0.03& 0.050\\ 
$4.79\times 10^4$& 5& $1.99$&  0.082& 0.357& 0.39& 2.11&
 0.99& 0.107\\ 
$7.59\times 10^4$& 5& $3.03$& 0.337& 0.179& 1.91& 0.59&
 0.16& 0.068\\ 
$8.99\times 10^4$& 5& $3.5$&0.391& 0.176& 2.02& 0.48&
0.14& 0.073\\ 
$1.12\times 10^5$&5&$ 4.3$&0.471& 0.148& 2.13& 0.37& 0.12& 0.057\\
$1.58\times 10^5$&5 &$5.99$&0.563& 0.142& 2.30& 0.20& 0.03& 0.043\\
$3.16\times 10^5$&5&$ 11.3$&0.658& 0.127& 2.40& 0.10&0.02& 0.047\\
$1.99\times 10^4$& 14.0& $2.5$&0.345& 0.195& 1.95& 0.55
&-0.07& -0.029\\ 
$3.16\times 10^4$& 9.2&$ 2.5$&0.323& 0.210& 1.85& 0.65
&0.03& -0.006\\ 
$3.98\times 10^4$& 7.45& $2.5$&0.266& 0.222& 1.64&
0.86&0.24& 0.051\\  
$7.94\times 10^4$& 3.95& $2.5$& 0.233& 0.255& 1.53&
0.97 &0.35& 0.084\\ 
$1.00\times 10^5$& 3.2&$ 2.5$&0.191& 0.257& 1.53&
0.97&0.35& 0.127\\ 
$3.16\times 10^5$& 1.1&$ 2.5$& 0.173& 0.289 &1.25&
1.25&0.63& 0.143\\ 
\hline
\multicolumn{9}{l}{All the systems have initially $W_0=7$ and
$\alpha_{i}=2.5$}\\ 
\multicolumn{9}{l}{Final values are calculated at $t=15$ Gyr}\\
\end{tabular}
\end{table}
\newpage
\clearpage
\begin{table}
\begin{tabular}{|cccccc|}
\multicolumn{5}{c}{\bf Table 3}\\
\hline
$id$&$N$&$ M_f/M_i$&$ M_{wd}/M_f$&$ \alpha_f$&$
t_{cc}$(Gyr)\\
\hline

I&4096&0.195& 0.273& 1.38& 6.0\\
II&4096&0.242&0.268& 1.50& 7.5\\
III&4096&0.195&0.281&1.44& 5.5\\
IV&4096&0.202&0.286& 1.32& 7.1\\

I&8192&0.127& 0.365& 0.86& 6.3\\
II&8192&0.156&0.325&1.25& 6.9\\

I&16384&0.075& 0.422& 0.54& 7.6\\
\hline
\multicolumn{5}{l}{Disk shocking is not included in these
simulations.}\\
\multicolumn{5}{l}{Final values are calculated at $t=15$ Gyr.}\\
\multicolumn{5}{l}{Initial conditions:}\\
\multicolumn{5}{l}{$W_0=5,~\alpha_i=2.5,~R_g=4
\hbox{Kpc},~M_i=6.15\times 10^4 M_{\odot}$}\\
\multicolumn{5}{l}{$t_{cc}$ is the time of core collapse}\\

\end{tabular}
\end{table}
\newpage
\clearpage
\section*{Figure captions}
Figure 1 Time evolution of the main properties of systems starting
with $\alpha_i=2.5$, $W_0=7$, $M_i=6.15 \times 10^4 \cmo$ and having
galactocentric distances equal 
to $R_g=4$ Kpc (solid line),  $R_g=5$ Kpc (dotted line),  $R_g=8$ Kpc
(short-dashed line),  $R_g=16$ Kpc (long-dashed line).
(a) Evolution of the total number of stars;(b) evolution of the ratio
of total mass to initial mass; (c) evolution of the total number of
stars having a mass in the range 
$0.1-0.12~m_{\odot}$ (upper curves) and $0.25-0.31~m_{\odot}$ (lower
curves) for the systems located at $R_g=4$ Kpc (solid lines) and at
$R_g=16$ Kpc (long-dashed lines); (d) evolution of the ratio of the
total mass of neutron stars  to the total mass at time $t$; (e)
evolution of the ratio of total mass of white dwarfs to the total 
mass at time $t$; (f) Evolution of the slope of the mass
function calculated for main sequence stars in the range
$0.1<m/m_{\odot}<0.5$.\\ 
Figure 2 Time evolution of the mass function for main sequence stars
for the system with $\alpha_i=2.5$, $W_0=7$,  $M_i=6.15 \times 10^4
\cmo$  and $R_g=4$ Kpc. 
The four curves shown correspond (from the upper to the lower one) to
$t=0$ Gyr, $t=7$ Gyr, $t=10$ Gyr, $t=15$ Gyr.\\
Figure 3 (a) Time evolution of the ratio of total mass to initial mass
for systems
with $\alpha_i=2.5$, $W_0=7$, $M_i=6.15 \times 10^4 \cmo$ and $R_g=4,5,8,16$ Kpc (symbols as in
Figure 1). Straight lines take into account only the mass loss by relaxation,
curved lines only the mass loss by stellar evolution. The latter curve
does not depend significantly on the galactocentric distance of the
system and the 
four curves are almost indistinguishable; (b) evolution of the ratio of total
mass to initial mass for systems
with $\alpha_i=2.5$, $W_0=7$, $M_i=6.15 \times 10^4 \cmo$ and
$R_g=4,5,8,16$ Kpc (symbols as in 
Figure 1) including only the mass loss by relaxation versus time
scaled by the value of $F_{cw}$ for each system.\\
Figure 4 (a) Fraction of the initial mass remaining in a cluster after
$15$ Gyr, $M(15\hbox{Gyr})/M_i$, as a function of $\log M_i$ and $R_g$
for systems with $\alpha_i=2.5$ and $W_0=7$; 
(b) contours of constant values of $M(15\hbox{Gyr})/M_i$, as indicated
on the right end of the 
curves shown in the plot, as a function of
the galactocentric distance and of the logarithm of the initial
mass.\\
Figure 5 Slope of the mass function at $t=15 $ Gyr measured for main
sequence stars with masses in the range $m=[0.1,0.5]~m_{\odot}$ from the
analytic expression derived in the text (see eq.(\ref{nemp})) as a function
of galactocentric distance for the following values of the initial
mass of the cluster (from the upper to the lower curve): $\log
M_i=6,5.5,5.2,5,4.79,4.5 $. Initial conditions are $\alpha_i=2.5$,
$W_0=7$. Dots represent values obtained from $N$-body simulations for
$\log M_i=4.79$.\\
Figure 6 Time evolution of the slope of the mass function of main
sequence stars with $0.1<m/m_{\odot}<0.5$ calculated in three
different 3-d and 2-d shells (see text for their exact limits): data
for the innermost shell are plotted by a solid line and circles, for
the intermediate shell by short-dashed line and triangles, for the
outermost shell by long-dashed line and crosses.\\
Initial conditions are $\alpha_i=2.5$,$W_0=7$, $M_i=6.15 \times 10^4
\cmo$ and galactocentric distance as indicated below.
(a) 3-d shells, $R_g=4 $ Kpc; (b) 2-d shells, $R_g=4 $ Kpc; (c) 3-d
shells, $R_g=16$ Kpc;  (d) 2-d shells, $R_g=16$ Kpc.\\
Figure 7 (a) Evolution of the difference between the slope of the
mass function in 3-d shells  and  the slope of the initial mass
function (all measured for main 
sequence stars with $0.1<m/m_{\odot}<0.5$)  for the system at $R_g=4$
Kpc. Initial conditions of the systems: $W_0=7,~\alpha_i=2.5,
M_i=6.15\times 10^4~\cmo$.
(b) Evolution of the difference between the slope of the
mass function in 3-d shells  and  the slope of the global mass
function at time $t$
(all measured for main 
sequence stars with $0.1<m/m_{\odot}<0.5$)  for the system at $R_g=4$ Kpc.\\
(c)-(d) same as (a) and (b) but  for the system at $R_g=16$ Kpc.
Symbols as in figure 6.\\
Figure 8 (a) Time evolution of the ratio of total mass in white dwarfs at time $t$ to the
total initial mass for systems with $\alpha_i=2.5$,
$W_0=7$, $M_i=6.15 \times 10^4~\cmo$ and $R_g=4$ Kpc (solid line), 5 Kpc (dotted line), 8 Kpc
(short-dashed line), 16 Kpc (long-dashed line). The dot-dashed line is
the fraction of total mass in white dwarfs 
expected if all the white dwarfs produced were retained in the cluster
and none escaped due to relaxation (see eq.(\ref{wdt}) in the text),
$M_{wd,prod}$.\\
(b) Ratio of the total mass of white dwarfs retained in a system at
time $t$ to the total mass of white dwarfs actually produced for
systems with $\alpha_i=2.5$,
$W_0=7$, $M_i=6.15 \times 10^4 \cmo$ and $R_g=4$ Kpc (solid line), 5
Kpc (dotted line), 8 Kpc 
(short-dashed line), 16 Kpc (long-dashed line) versus time.\\
(c) Same as (b) but with time scaled by the parameter $F_{cw}$. Data
from runs with $W_0=5$ are also shown in this plot. (symbols
distinguishing different $R_g$ as in
(b)).\\
(d)  Ratio of total mass in white dwarfs at time $t$ to the
total mass of  white dwarfs produced at time $t$ versus $t/F_{cw}$
(average values of the two runs starting 
$W_0=5$ and $W_0=7$ both with $\alpha_i=2.5$, $R_g=4$ Kpc, $M_i=6.15
\times 10^4 \cmo$). The line
superimposed is the result of a polynomial fit (see eq.(\ref{polfit}) in the
text).\\
Figure 9 Fraction of the total mass at $t=15$ Gyr in white dwarfs as a
function of the parameter $F_{cw}$ calculated by eq.(\ref{wdr}) (see text).\\
Figure 10  Fraction of the total mass at $t=15$ Gyr in white dwarfs as a
function of $\log M(15~\hbox{Gyr})$ at different distances from the
galactic center (calculated by eq.(\ref{wdr}) in the  text).\\
Figure 11 In all figures dashed lines  show the time evolution of the
slope of the mass 
function for clusters with (from the lower to the upper curve)
$F_{cw}=20000,25000,40000,80000$ calculated by eq.(\ref{nemp}) in the text.\\
(a) Contours of equal values of $M_{wd}(t)/M(t)$ as a function
of time, $t$, and slope of the mass function at time $t$ calculated
according to eq.(\ref{wdr}) in the text taking $\alpha_i=2.5$ and $W_0=7$.\\
(b) Contours of equal values of $M_{wd}(t)/M(t)$ as a function
of time, $t$, and slope of the mass function at time $t$
calculated by 
extrapolating the current properties of a cluster back in time
(``observational'' procedure; see text for further details).
(c)Contours of equal values of the ratio of the theoretical estimate
of  $M_{wd}(t)/M(t)$ to the 
``observational'' one as a function
of time, $t$, and slope of the mass function at time $t$.\\
Figure 12 As in figure 11 but as a function of the logarithm of the
mass of the cluster at $t=15$ Gyr and of its galactocentric distance.\\
Figure 13  Same as figure 5 with arrows showing the additional change
in the slope of the mass function due to the effects of disk shocking
(data from $N$-body simulations).\\
Figure 14 (a) Comparison of the time evolution of $M(t)/M_i$ for
systems with the effects of disk shocking and without them. All the
systems start with $\alpha_i=2.5$ and $W_0=7$, $M_i=6.15 \times 10^4
\cmo$; galactocentric
distances are $R_g=4,5,8,16$ Kpc (solid,dotted,short-dashed and
long-dashed line respectively).For each pair the lower curve is the
one relative to the run with disk shocking, with the exception of the runs at
$R_g=16$ Kpc for which the lower one refers to the run without disk
shocking.\\
(b) Same as (a) for the evolution of the slope of the mass
function.\\
Figure 15 Fraction of the total mass at $t=15$ Gyr in white dwarfs as
a function of the fraction of the  initial mass left in the cluster at
$t=15$ Gyr. Solid line is calculated by the analytical expression
(eq.(\ref{wdr})) derived in the text. Dots are data from $N$-body
simulations with $\alpha_i=2.5$ (full dots from runs without disk
shocking, see Table 1 and Table 3; 
circles from runs with disk shocking, see Table 2; triangles from runs
starting with $N=8192$ and the cross refers to the run with
$N=16384$ see Table 3).\\
Figure 16 Difference between the initial value of the  slope
of the mass function ($\alpha_i=2.5$) and  its value at $t=15$ Gyr as
a function of the fraction of the  initial mass left in the cluster at
$t=15$ Gyr. Solid line is calculated by the analytical expression
(eq.(\ref{nemp})) derived in the text. Dots are data from $N$-body
simulations (symbols as in figure 15).\\
Figure 17 Time evolution of the ratio of mass at time $t$ to initial
mass not considering  mass loss due to stellar evolution for four runs with
disk shocking. Straight lines show the best linear fit of the curves
obtained from $N$-body data. From the lower to the upper one , the
curves refer to the following initial conditions 
$\log M_i=4.68,~R_g=5~\hbox{Kpc}$ ,
$\log M_i=5.5,~R_g=1.1~\hbox{Kpc}$, 
$\log M_i=4.3,~R_g=14~\hbox{Kpc}$, 
$\log M_i=5.5,~R_g=5~\hbox{Kpc}$.\\ 
Figure 18 Slope of $M(t)/M_i$ (not taking into account mass loss by stellar
evolution)  for runs with disk shocking from $N$-body data versus the
analytical expression derived in the text (see eq.(\ref{slp})). The
line is given by $\lambda(\hbox{N-body})=\lambda(\hbox{fit})$.\\
Figure 19 (a) $M(15 \hbox{Gyr})/M_i$ versus the parameter $F_{cw}$ for
runs with (circles) and without (full dots) disk shocking. Solid
line is calculated by the analytical expression derived in the text
(eq.(\ref{mlo})). (b) Same as (a) but with the family parameter $F_{cw}$ replaced by
the equivalent family parameter $F_{cw}^{eq}$ defined in the text.\\
Figure 20 Contours of equal values (given at the right side of each
pair of curves)  of the family parameter $F_{cw}$
(solid lines) and of the equivalent family parameter, $F_{cw}^{eq}$,
defined in the text.\\
Figure 21 Contours of equal values of the mass of a cluster at $t=15$
Gyr with (dashed lines) and without (solid lines) the effects of disk
shocking; initial conditions $W_0=7$ and $\alpha_i=2.5$. The number
beside each curve gives $\log M(15 \hbox{Gyr})$.\\
Figure 22 Slope of the mass function at $t=15 $ Gyr for main
sequence stars with masses in the range $m=[0.1,0.5]~m_{\odot}$
without (solid lines) and with (dashed lines) the effects of disk
shocking calculated by the
analytic expression derived in the text (eq.(\ref{nemp}) with $F_{cw}$
replaced by $F_{cw}^{eq}$ for the case with disk shocking) as a function
of galactocentric distance for the following values of the initial
mass of the cluster (from the upper to the lower curve): $\log
M_i=6,5.5,5.2,5,4.79,4.5 $. Initial conditions are $\alpha_i=2.5$,
$W_0=7$. \\
Figure 23 Slope of $M(t/F_{cw})/M_i$ for systems starting with
$W_0=5,~R_g=4~\hbox{Kpc}, \alpha_i=2.5$ and $M_i=6.15 \times 10^4
\cmo$ versus the initial number of 
stars in the simulation $N$. Disk shocking was not included in these
runs. Four runs have been done for $N=4096$ and two runs for $N=8192$.\\
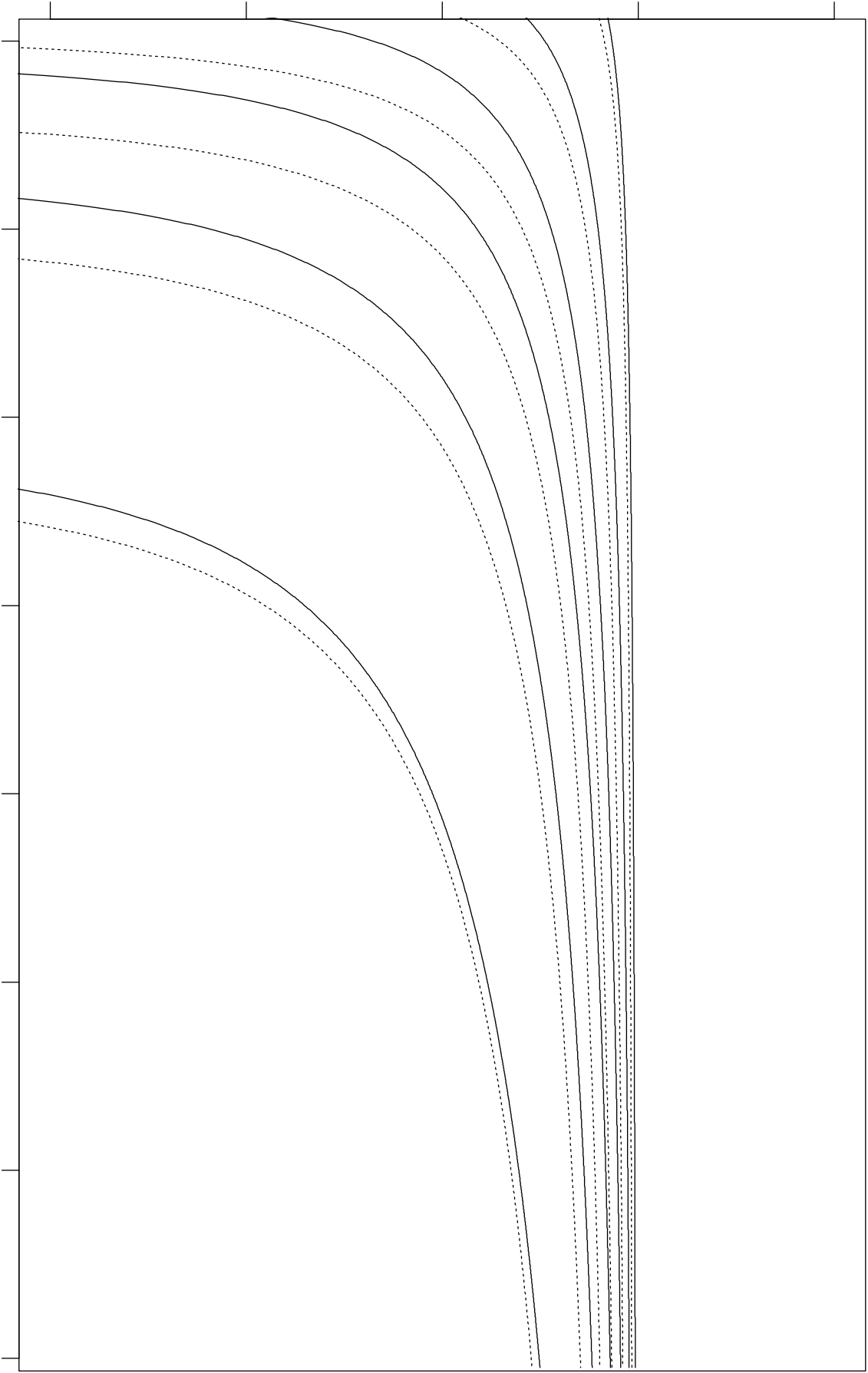
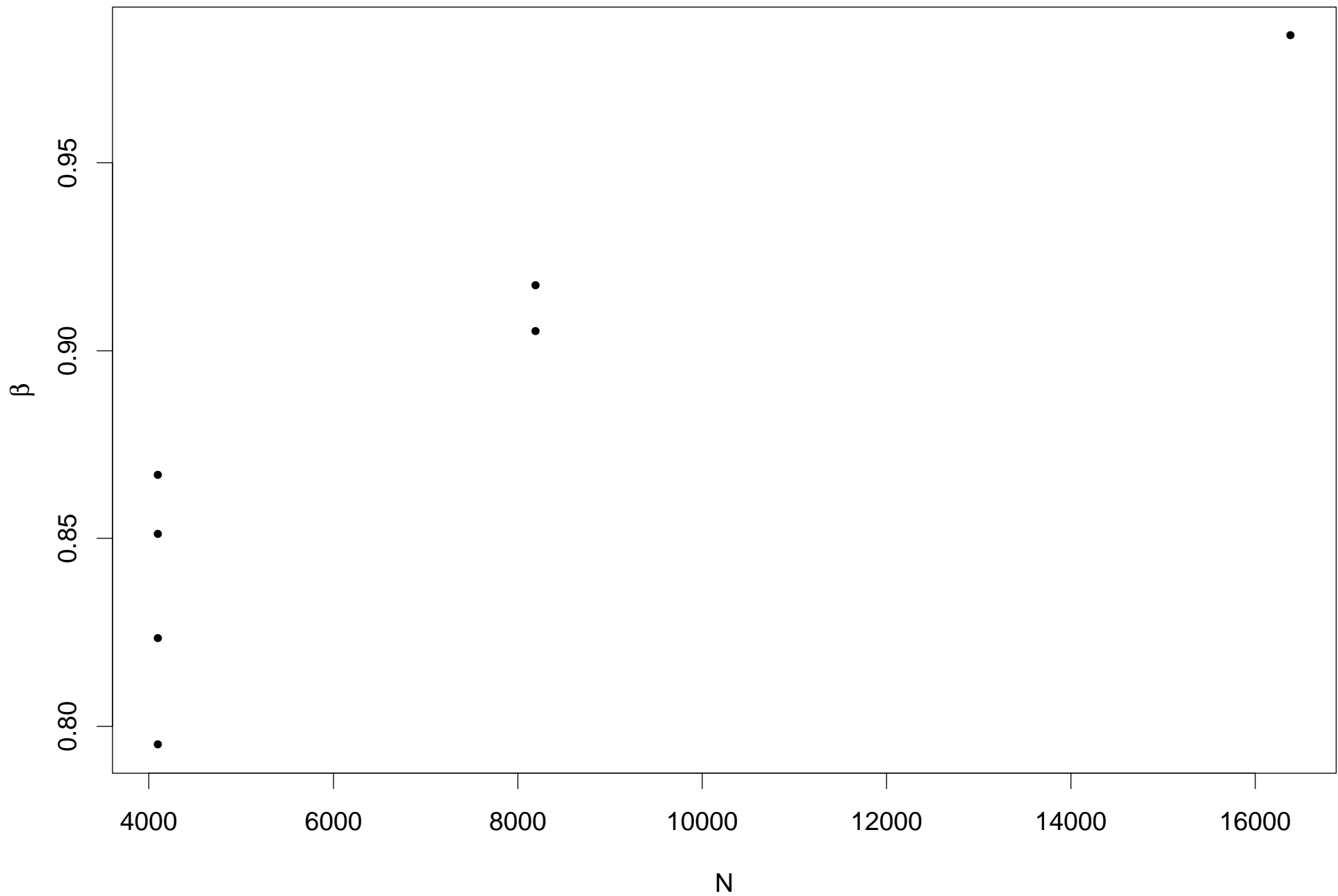
\end{document}